\documentclass[pre,twocolumn,showpacs,amsmath,amssymb]{revtex4}
\usepackage{graphicx}
\usepackage{bm}
\usepackage{natbib}

\begin{document}

\title{The dynamics of gene duplication and transposons in microbial genomes following a
sudden environmental change}

\author{Nicholas Chia}
\email{chian@uiuc.edu}

\author{Nigel Goldenfeld}

\affiliation{Institute for Genomic Biology, University of Illinois
at Urbana-Champaign, 1206 West Gregory Drive, Urbana, IL 61801}

\affiliation{Loomis Laboratory of Physics, University of Illinois at
Urbana-Champaign, 1110 West Green Street, Urbana, IL 61801}

\date{\today}

\begin{abstract}

A variety of genome transformations can occur as a microbial population
adapts to a large environmental change.  In particular, genomic surveys
indicate that, following the transition to an obligate, host-dependent
symbiont, the density of transposons first rises, then subsequently
declines over evolutionary time. Here, we show that these observations
can be accounted for by a class of generic stochastic models for the
evolution of genomes in the presence of continuous selection and gene
duplication.  The models use a fitness function that allows for partial
contributions from multiple gene copies, is an increasing but bounded
function of copy number, and is optimal for one fully adapted gene
copy.  We use Monte Carlo simulation to show that the dynamics result
in an initial rise in gene copy number followed by a subsequent
fall-off due to adaptation to the new environmental parameters. These
results are robust for reasonable gene duplication and mutation
parameters when adapting to a novel target sequence. Our model provides
a generic explanation for the dynamics of microbial transposon density
following a large environmental changes such as host restriction.

\end{abstract}

\pacs{87.10.-e, 87.10.Mn, 87.23.Kg}


\maketitle

\section{Introduction}

Biological evolution involves a variety of genome transformations,
which include point mutation, homologous recombination, gene
duplication, and horizontal gene
transfer~\cite{syvanen1994hgt,taylor2004duplication,lynch2007origins},
and which involve a variety of mobile genetic
elements\cite{smalla2002pad,OSBO2002,Frost05}. Understanding the
influence that each of these mechanisms and elements exerts on the
process of evolution is one of the current frontiers in
biology~\cite{goldenfeld2007cbs}. Transposons are one such genetic
element---capable of copying and pasting segments from one location to
another within a genome, they provide vehicles for gene
duplication~\cite{kleckner1981transposable,feschotte2007dna}.

Transposons are copied and inserted across genomes through a variety of
mechanisms~\cite{kleckner1981transposable,feschotte2007dna}.
Non-conservative transposons multiply within a genome by replicating
themselves elsewhere. Many code for the proteins that copy and insert
themselves throughout the genome. Other transposable elements are more
passive, relying on proteins from other transposons in order to
proliferate.

Insertion sequences, or IS elements, are a particular type of
non-conservative transposon, which contain their own proteins for
replication but typically do not contain any additional
proteins~\cite{mahillon1998insertion,filee2007insertion}. However, when
two IS elements are near each other along a genome---for example,
flanking both sides of a gene---they may form a composite transposon
that includes the host genes sandwiched between the IS
elements~\cite{kleckner1981transposable,mahillon1998insertion}. These
composite transposons have been associated with the evolution of
virulence factors that affect infection and severity of
diseases~\cite{quintiliani1996characterization,doublet2009association},
and have been implicated in large chromosomal
rearrangements~\cite{watanabe2007precise}.

Due to the conserved tendency of the genes required to replicate and
insert an IS element, it is possible to estimate the IS element density
within a genome through sequence analysis.
In order to probe the evolutionary dynamics of microbial genomes, Moran
and Plague estimated the IS density in Bacteria following host-restriction,
the transition to becoming an obligate, host-dependent, organism such as a
gastrointestinal symbiont~\cite{moran2004genomic}.
They found that on average initial host restriction was followed by a sharp
increase in IS density but that at long times the IS density generally
declined to near zero. This overall pattern has been more directly validated
in a few taxonomic lineages by tracking the evolution of particular closely
related genomes~\cite{moran2003tracing,plague2008extensive}. This pattern may
be sufficiently widespread, and if so, it should have a generic explanation,
independent of the particular organisms or environment, and this is what we
seek to provide in the present article.

In order to explain the trend in IS density following host-restriction,
we focus on the role of IS elements as vectors for gene amplification
through their roles in composite
transposons~\cite{kleckner1981transposable,mahillon1998insertion}.
Several studies have indicated the importance of gene amplification in
the rapid evolution of
Bacteria~\cite{andersson1998evidence,hendrickson2002amplification}.
Duplicated genes provide a basis for the evolution of novel
function~\cite{hughes1994evolution,bergthorsson2007osd}, and have been
implicated in the evolution of new organismal
forms~\cite{ohno1970evolution} and lineages~\cite{serres2009evolution}.
Gene duplication events have been invoked in medically important traits
and diseases in humans~\cite{conrad2007gene}, including various forms
of cancer~\cite{turner1988mutations,ciullo2002initiation}, as well as
in the expansion of gene families such as the
globins~\cite{higgs1989review} and the DNA replication processivity
complex subunits~\cite{chia2010dna}. Even whole genome duplications
have been observed in many organisms including
yeast~\cite{kellis2004proof}, small flowering
plants~\cite{vision2000origins}, and
pufferfish~\cite{christoffels2004fugu,jaillon2004genome}.

If gene duplications confer an adaptive advantage to their hosts, then
one might expect a concomitant proliferation in the IS elements that
provide the mechanism for gene duplication---especially when adapting
to a new environment. Conversely, in relatively consistent environments
gene duplications confer no advantage to a well-adapted organism and we
might anticipate selection pressure for a decrease in IS density. Thus,
our strategy for interpreting the genomic trends reported by Moran and
Plague~\cite{moran2004genomic} is to better understand the evolutionary
dynamics of gene duplication, and thence to infer the corresponding
dynamics of the IS elements.

\begin{figure}[t]
\begin{center}
\includegraphics[width=0.9\columnwidth]{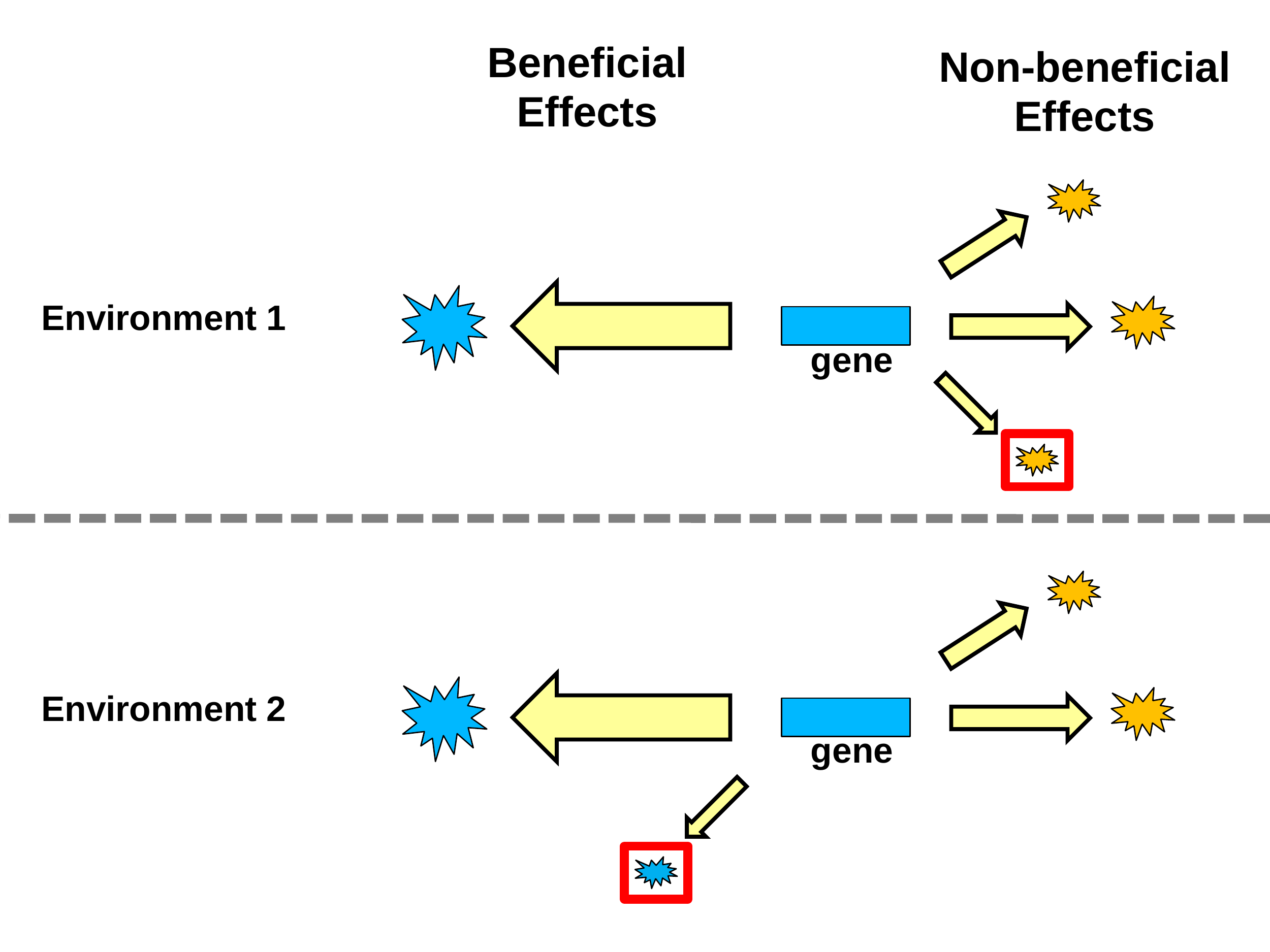}
\caption{\label{fig:side-effect} (Color online). Schematic: Gene activity
following a change in environmental conditions. Each gene gives rise to
a gene product that has a number of multiple activities that effect
organismal fitness differently. Some of these activities are
beneficial, while others are deleterious. In general, the selected
genes tend to maximize the benefit while minimizing any harmful
``side-effect''. However, when an organism undergoes a change in its
environmental conditions, previously deleterious activities may become
beneficial and vice-versa.}
\end{center}
\end{figure}

In order to probe the link between gene duplication and adaptation to a
novel environment, we model genome dynamics with gene duplication.
Previous evolutionary modeling has largely focused on the process of
mutation and recombination~\cite{kimura1971theoretical,desai2007speed},
large scale genome
duplications~\cite{maere2005modeling,roth2007evolution}, or static
features such as copy number distribution~\cite{yanai2000predictions}.
Here we quantify a continuous selection mechanism for the evolution of
novel genes put forth by Bergthorsson \textit{et
al}~\cite{bergthorsson2007osd}. In our model, we consider a protein
encoded by a gene that has multiple activities---for example, enzymatic
activity or non-specific binding. As shown schematically in
Fig.~\ref{fig:side-effect}, the initially deleterious effect of a gene
product may subsequently become beneficial to the organism when exposed
to different environmental conditions, as documented, for example, in
the growing body of literature on non-specific interactions involving
proteins~\cite{nobeli2009protein}.

There are then two mechanisms by which an organism may increase a
particular protein activity---efficiency and expression. In the first
mechanism, a gene may undergo mutations that result in a more effective
protein, i.e., a gene with some small activity which is favorable to
the organism can undergo mutations that subsequently enhance that
activity. In the second mechanism, increasing the level of gene
expression---for example, by creating additional copies of a gene
within the genome---also positively impacts the total activity of a
given gene product.

\begin{figure}[t]
\begin{center}
\includegraphics[width=0.9\columnwidth]{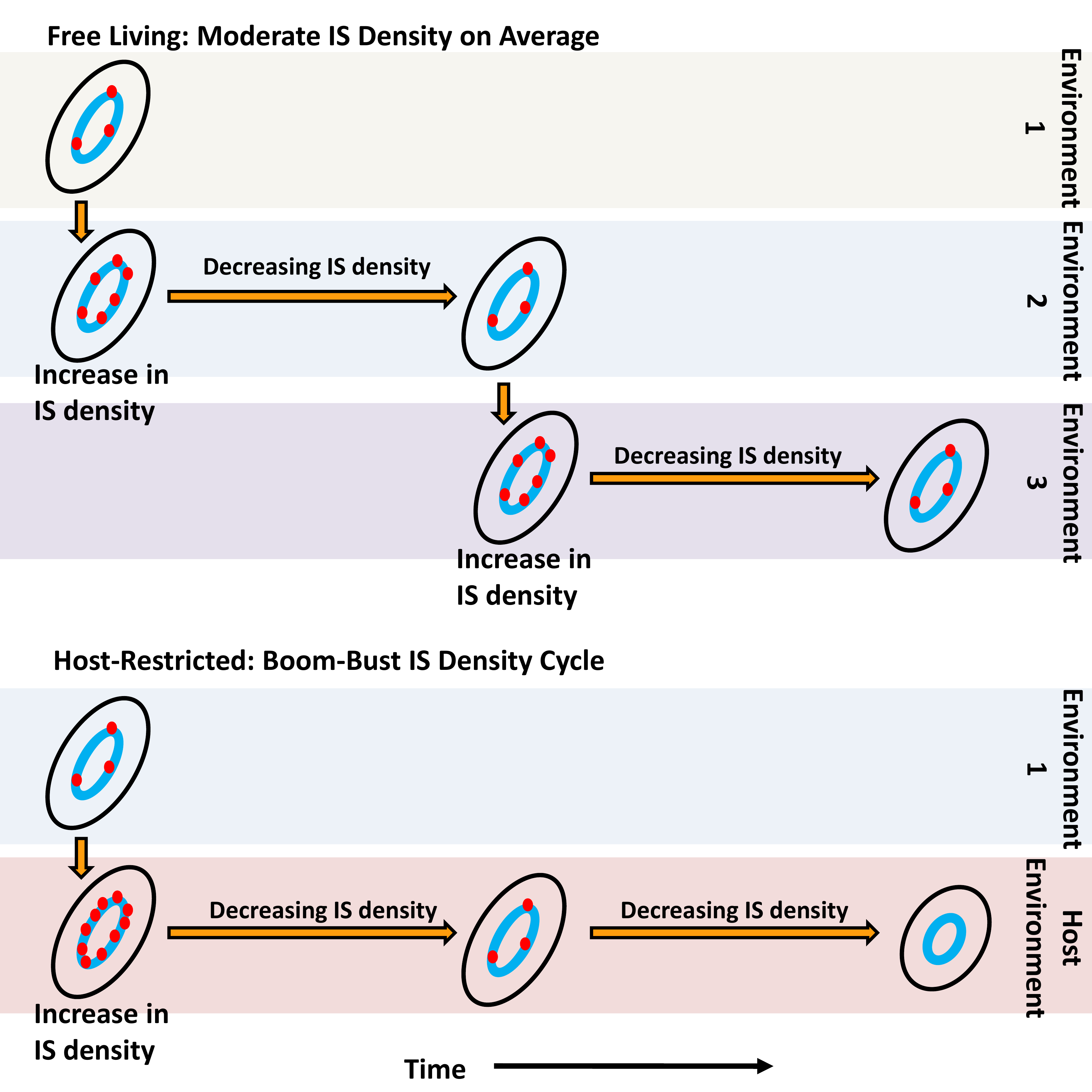}
\caption{\label{fig:bigdiagram} (Color online). Schematic: Adaptation to a
changing environment. We consider organisms experiencing a large change
in environmental conditions. Due to the role of gene duplications in
adapting novel functions, these changed conditions result in an
enhancement of the IS element density within the genome. Free-living
bacteria experience a fluctuating environment, which results in the
maintenance of IS density. Host-restriction represents a large change
in environmental conditions, resulting in an initial boom in IS
density, such as that seen in the recent obligate organisms. However, a
host also represents a stable and consistent environment. The lack of
any need for rapid adaptation leads to near-zero IS density by a
process of slow decay.}
\end{center}
\end{figure}

The purpose of this work is to show that adaptation to a large
environmental change provides a sufficient explanation for both the
short term increase and the long term decrease in IS density following
host restriction---as outlined in Fig.~\ref{fig:bigdiagram}. The rapid
adaptation to a new environment results in a large number of gene
duplications, presumably involving IS elements. Likewise, the long term
consistency of the host environment leads to a decreased number of
duplicates and near-zero IS density. We build a quantitative model of
the mechanism for adaptation via gene
duplication~\cite{bergthorsson2007osd}, and show that this model
accounts for the gross characteristics in IS density following host
restriction~\cite{moran2004genomic}.

This paper is organized as follows.  In Section~\ref{sec:model}, we
present our quantitative model of Bergthorsson \textit{et al.}'s
proposed mechanism for the emergence of novel genes under continuous
selection~\cite{bergthorsson2007osd}. Results of our simulations are
presented in section~\ref{sec:results}. Finally, in
section~\ref{sec:diss}, we describe the biological interpretation of
our work, showing how our results are consistent with the trends
observed by Moran and Plague\cite{moran2004genomic}.



\section{Model of Continuous Selection with Gene Duplication}~\label{sec:model}

We consider a population of replicating cells whose replication rates
depend on the genes within their genomes. We express the probability of
the $k$th cell replicating according to a fitness function
$\mathcal{F}$.
\begin{center}
\begin{equation}
\mathcal{F}_k = \min\left(\sum_{j=1}^n g_j,1\right) -
n\mu + \eta_d \label{eq:F1}
\end{equation}
\end{center}
where $n$ represents the number of total genes in the genome, $\mu$
denotes the fitness penalty arising from each additional gene, and
$g_j$ denotes the positive fitness contribution of the $j$th
gene within the genome. The parameter $\eta$ represents an assumed
additive Gaussian noise with standard deviation $d$. Introducing additive noise
allows us to vary the strength with which the organismal fitness is
coupled to the genome, allowing us to take into account the effect of
environmental fluctuations as well as mutation in the parts of the
genome that are unrepresented in this simple model. Note that the decoupling
noise we have introduced here differs from the demographic noise intrinsic to
a population~\cite{butler2009robust}. In section~\ref{sec:noise},
we will discuss the functional form of the noise in more detail. The value of $g$
depends inversely on the Hamming distance $d$ between the gene sequence
$\mathcal{S_j}$ and some target sequence $\mathcal{T}$, i.e.,
\begin{center}
\begin{equation}
g_j = 1 - \frac{d(\mathcal{S_j},\mathcal{T})}{N} = 1 - \frac{\sum_{\imath=1}^N
\delta(S_{j\imath},T_\imath)}{N}\label{eq:g}
\end{equation}
\end{center}
where $N$ represents the number of letters in each sequence (both
$\mathcal{S_j}$ and $\mathcal{T}$), $d(\mathcal{S_j},\mathcal{T})$
represents the Hamming distance between $\mathcal{S_j}$ and
$\mathcal{T}$, and $S_{j\imath}$ and $T_\imath$ represent the
$\imath$th letter of $\mathcal{S_j}$ and $\mathcal{T}$, respectively.


These parameters are intended to caricature a biological process
whereby a novel beneficial functionality captured by target sequence
$\mathcal{T}$ would provide an overall improvement in fitness given
by $g_j(\mathcal{Sj}\!=\!\mathcal{T})=1$. Partially matching sequences
also provide some of the catalytic activity necessary for the new
function, yielding a partial benefit in proportion to the homology
between the gene sequence $\mathcal{S_j}$ and the target sequence
$\mathcal{T}$, in accordance with the continuous selection model of
Bergthorsson {et al.}~\cite{bergthorsson2007osd}. The total benefit
of a set of genes must also be less than the maximum benefit arising
from a single gene. Therefore, each gene has a fixed
cost $\mu$ that represents the deleterious effect of non-specific
interactions of the product coded for by a gene. For simplicity, we assume
that each gene in the genome is expressed equally without regard to the
more detailed considerations of gene regulation. This simplification
has its basis in recent work that has indicated that gene copy number
is positively correlated with gene expression
level~\cite{mileyko2008small}.

The scheme outlined above clearly sets the optimum at a the one gene solution,
with the gene matches the target sequences. This assumption comes from
the biological considerations of the need to minimize the deleterious
non-specific interactions while reaching a certain level of functional activity
for a given biomolecule. Note that while the optimum is chosen by design,
the behavior of the system as it evolves in time is not. We do not \textit{a priori}
know whether the optimum behavior will be to duplicate or to not duplicate. After
all, the long term advantage is for the single gene case, and it might be
superfluous to duplicate genes only the then try to reduce their number. Furthermore,
if we introduce high gene duplication rates, will competition suffice to overcome the
duplication rate and drive the reduction in the number of genes? Or will the
behavior of the system depend on the careful tuning of these parameters?
Constructing a simple model enables us to answer these questions and to probe the
viability of the scheme we outline in Fig.~\ref{fig:bigdiagram}.

\begin{figure}[t]
\begin{center}
\includegraphics[width=0.9\columnwidth]{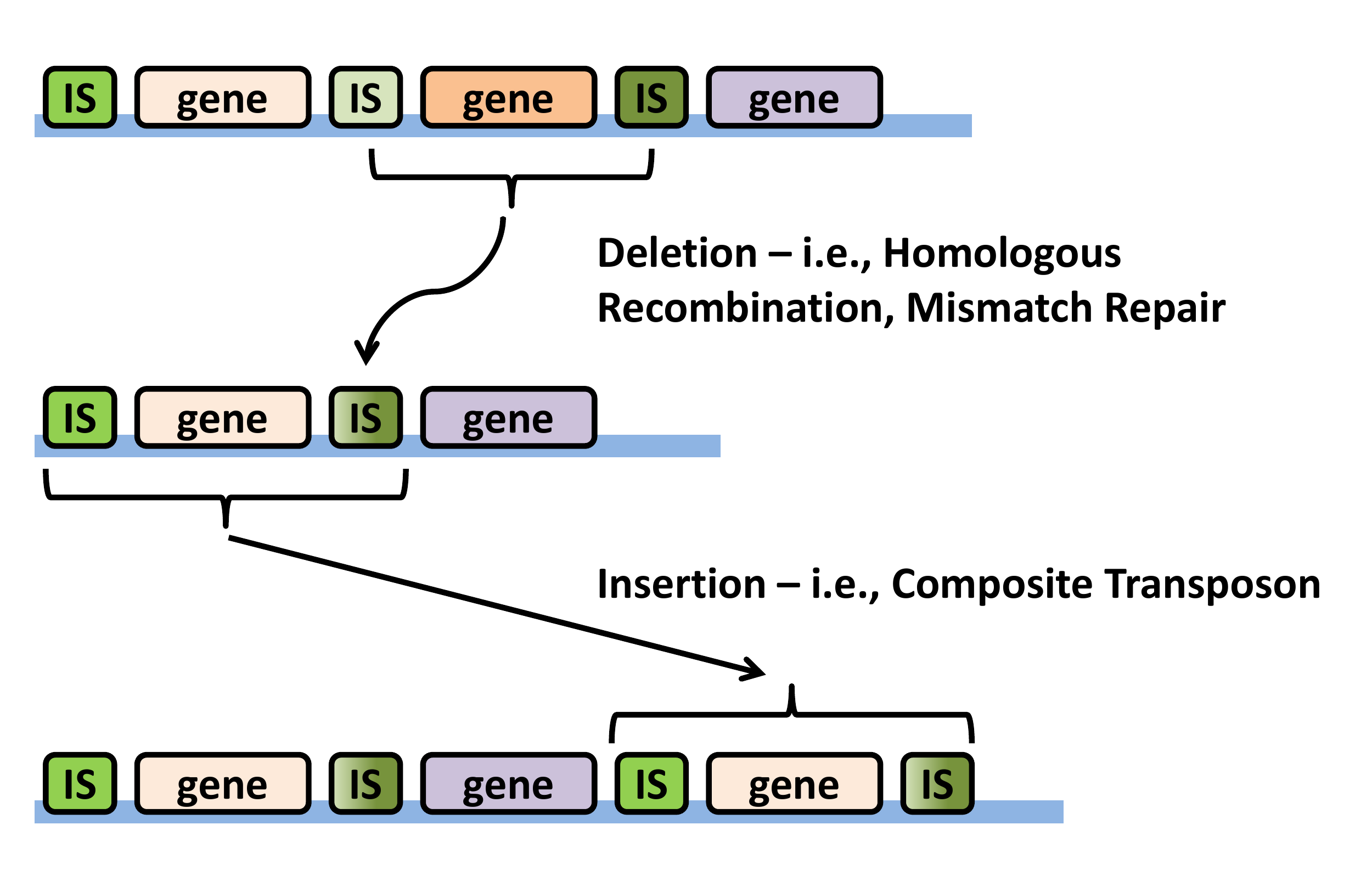}
\caption{\label{fig:dupl-del} (Color online). Schematic: gene
deletion and duplication. (Upper) Deletion can occur via homologous
recombination or mismatch repair between nearly identical IS elements
by using the IS elements as templates for homologous recombination.
(Lower) Gene duplication occurs when a composite transposon that is
made up of two flanking IS elements replicates and inserts itself in a
different part of the genome.}
\end{center}
\end{figure}

Genes are allowed to evolve by spontaneous point mutation, internal
duplications, or deletions.
Spontaneous mutation occurs by replacing a letter at a particular
position within a gene, chosen at random from a uniform distribution,
and replacing that letter with a randomly chosen one from an alphabet
of size $c$. Gene duplications and deletions, just as with spontaneous
mutations, occur on randomly chosen genes within the population.
Duplications are modeled as insertion events, and do not overwrite
existing genes. Fig.~\ref{fig:dupl-del} outlines the processes that lead
to gene duplication and deletion.

Initially, all organisms contain a single identical copy of a randomly
chosen gene. At each time $t$, the fitness of a random cell $k$ is selected out
of a fixed population size and its fitness $\mathcal{F}_k$ from
Eq.~\ref{eq:F1} is measured. This number then defines the probability
that this cell will replicate at time $t$. If it is decided that the
organism replicates, it then overwrites a different randomly chosen
organism in the population---the so-called roulette scheme~\cite{goldberg1991comparative}.
Thus, organisms are on average being
selected for higher replication rates (defined by $\mathcal{F}$). One
generation is defined as the time it takes for half the total
population to undergo a growth attempt. A random update scheme governs
which organisms will attempt to replicate.

\section{Simulation Results}~\label{sec:results}

\begin{figure}[t]
\begin{center}
\includegraphics[width=0.9\columnwidth]{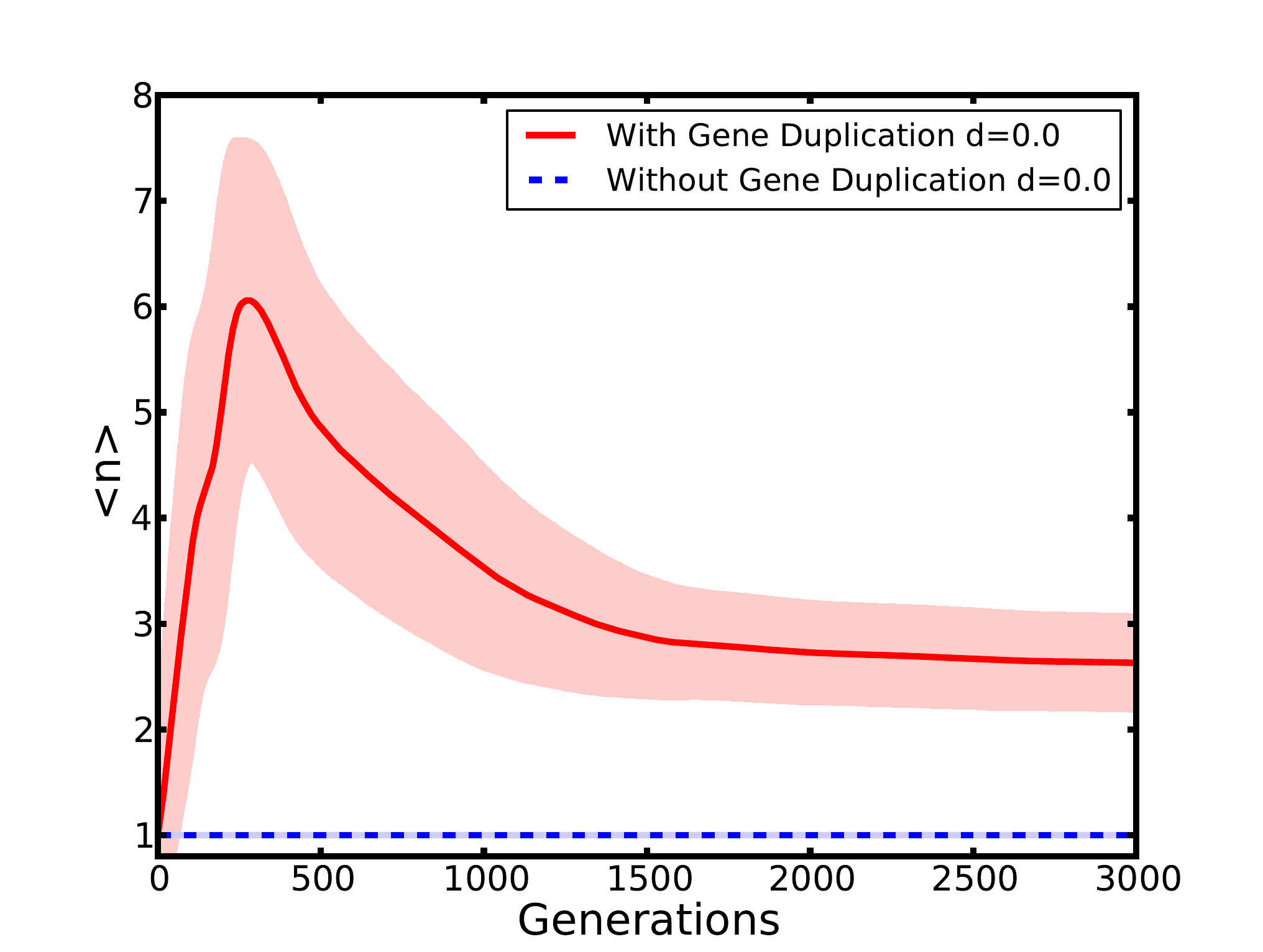}
\caption{\label{fig:gene_num} (Color online). Average gene number
$\langle n \rangle$ with and without gene duplication for simulations
with the fitness function given by Eq.~\ref{eq:F1}. The average is
given by the darker lines, while the lightly shaded areas represent
the area covered by the average standard deviation. In the initial
phases of adaptation, gene duplication dominates as the primary mode
for enhancing average fitness. As time passes, the slower mode of
adaptation provided by sequence mutation refines the genes and the
average number of genes per genome decreases. In the case with no gene
duplication or deletion the gene number remains constant. Simulations
were carried out with a duplication rate of $1$ per generation, a gene
deletion rate of 0.2 per gene per generation, a mutation rate of 0.01
per gene per generation, $N=10$, $c=10$, $\mu=0.05$, and $d=0.2$. We
considered a population of $10000$ organisms and averaged across $100$
replicate simulations with the same parameters but different initial
seeds. In the case without gene duplication, we set the gene
duplication and deletion rate to $0$ without change to the other
parameters.}
\end{center}
\end{figure}

We characterize the model of continuous selection with gene duplication
described in Section~\ref{sec:model} and contrast it with a model of
continuous selection with no gene duplication. In order to do so, we
assign a randomly chosen target sequence $\mathcal{T}$ and a randomly
chosen gene sequence that is initially fixed in the population (i.e.,
zero diversity at $t=0$). We then allow the system to evolve under
a selection pressure described by Eq.~\eqref{eq:F1}.

As shown in Fig.~\ref{fig:gene_num}, the average number of genes in a
genome $\langle n \rangle$ increases sharply at the onset through gene duplication.
As time passes and the individual genes become better adapted toward
the target sequence, the number of genes then begins to decrease. In
the long time limit, the gene number seemingly asymptotes to a small
number greater than $1$.

\begin{figure}[t]
\begin{center}
\includegraphics[width=0.9\columnwidth]{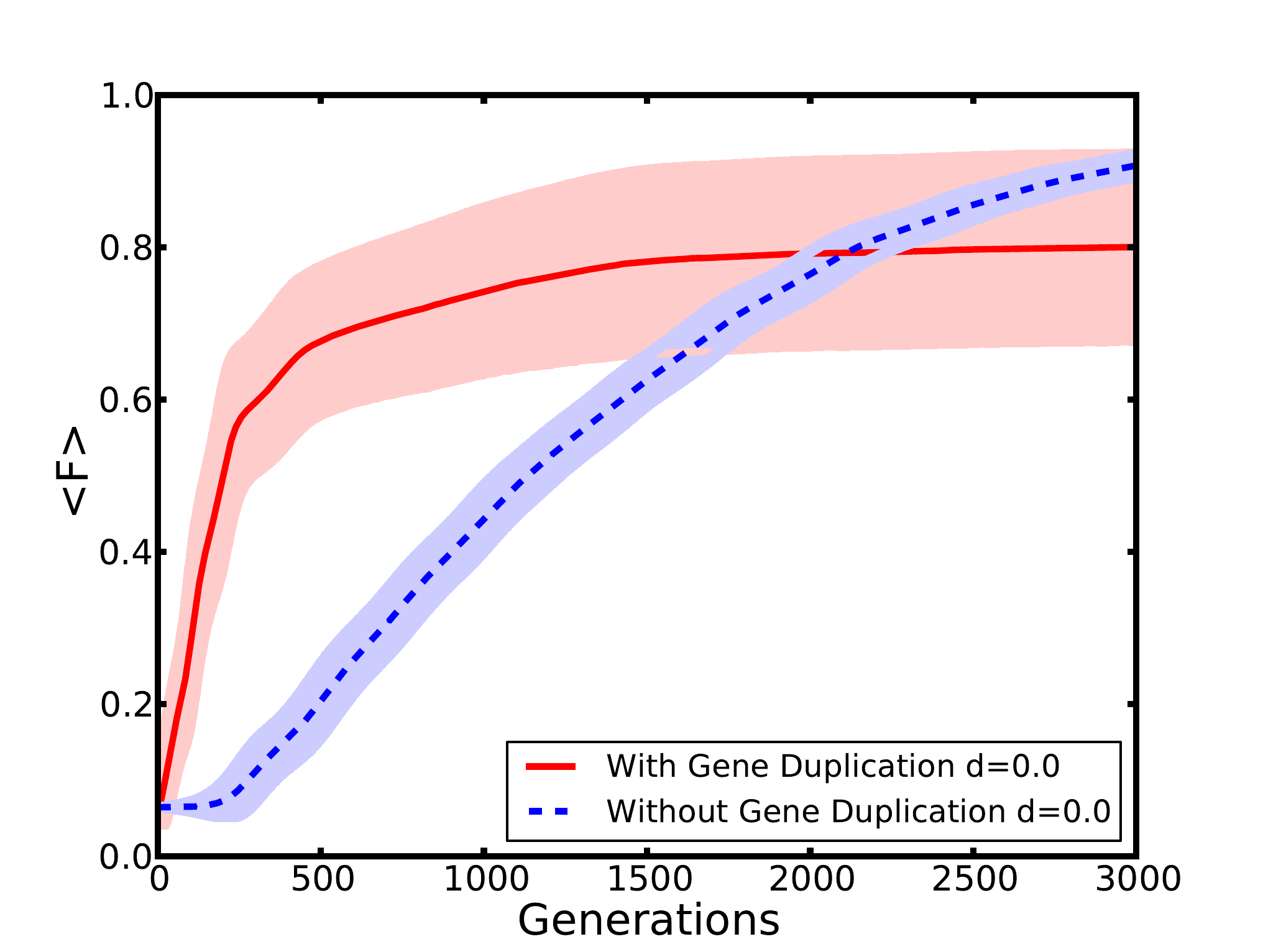}
\caption{\label{fig:org_fit} (Color online).  Average
organismal fitness $\langle \mathcal{F} \rangle$ with and without gene duplication.
With gene duplication the initial rate of adaptation is faster then in
the single gene case. However, at long times the single gene case
results in genes that are closer to the target sequence $\mathcal{T}$.
Parameters are the same as given in Fig.~\ref{fig:gene_num}.}
\end{center}
\end{figure}

These changes in gene number are accompanied by changes in the
individual gene scores or gene fitness $g$. Fig.~\ref{fig:org_fit}
plots the average organismal fitness $\langle \mathcal{F} \rangle$ as a function of
time and confirms that gene duplication is enhancing the initial rate
of adaptation (fitness increase) by providing a means for the organism
to amplify the benefit of a gene.

\begin{figure}[t]
\begin{center}
\includegraphics[width=0.9\columnwidth]{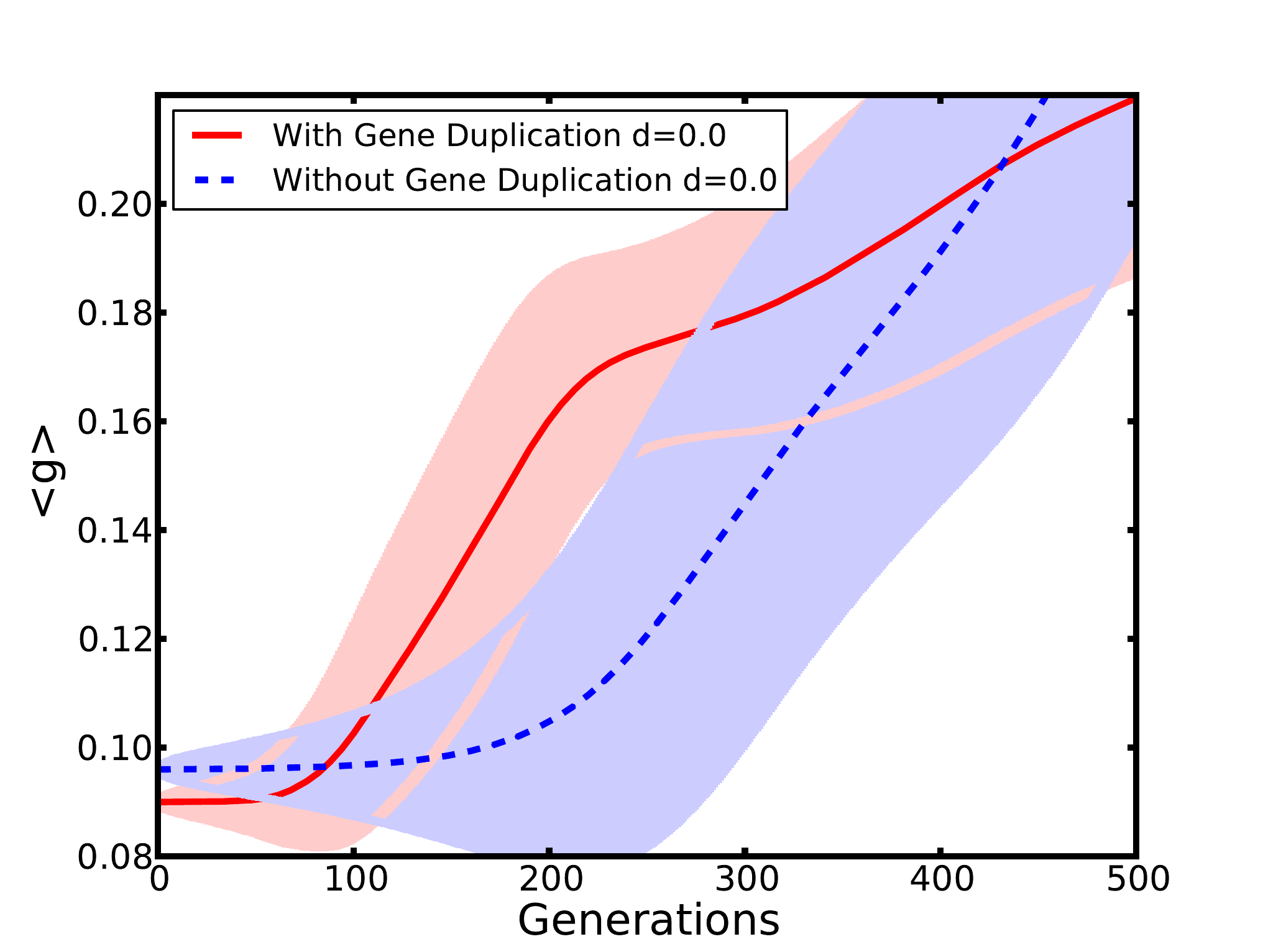}
\caption{\label{fig:gene_fit} (Color online).  Average gene fitness
$\langle g \rangle$ with and without gene duplication for simulations
with the fitness function given by Eq.~\ref{eq:F1}. With gene
duplication the initial rate of gene adaptation is faster than in the
single gene case. However, at longer times the single gene case results
in genes that are closer to the target sequence $\mathcal{T}$.
Parameters are the same as given in Fig.~\ref{fig:gene_num}. Note that
the horizontal axis scale differs from that shown in
Fig.~\ref{fig:gene_num}.}
\end{center}
\end{figure}

In contrast to the case of organismal fitness, average gene score or fitness
$\langle g \rangle$ does not necessarily increase with gene number $N$. In principle,
there is a tension between the greater mutation rate that larger copy
number engenders and the lesser effect on fitness from an individual
gene. By plotting the average gene score $\langle g \rangle$ in
Fig.~\ref{fig:gene_fit}, we see that, on average, genes initially adapt
faster toward the target sequence $\mathcal{T}$ with gene duplication.
It thus appears that the primary effect comes from mutation of the gene
copies provided by gene duplication, which leads to additional diversity
in comparison to the case without gene duplication.

\begin{figure}[t]
\begin{center}
\includegraphics[width=0.9\columnwidth]{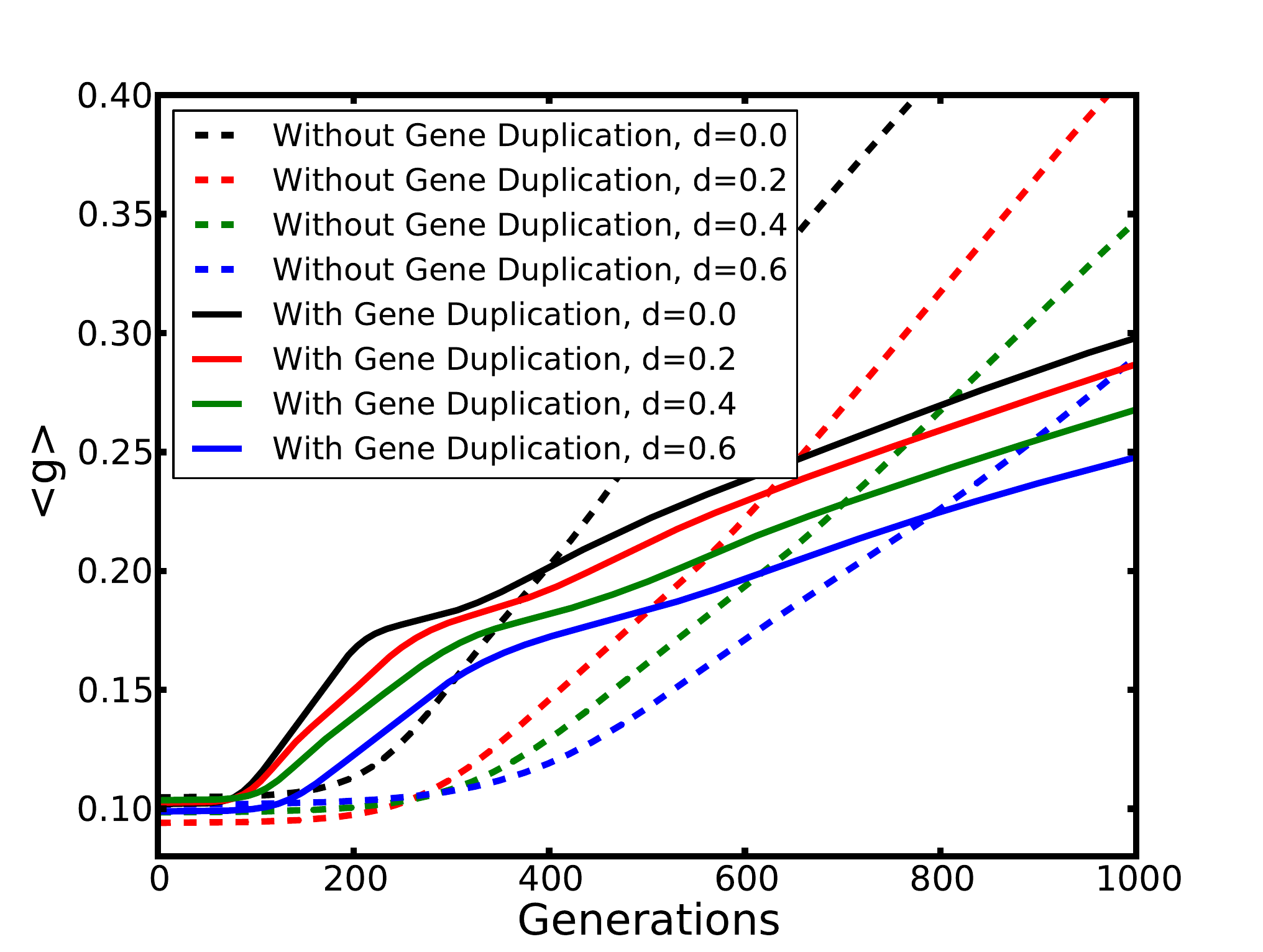}
\caption{\label{fig:gene_fit_noise} (Color online).  Average
gene fitness $\langle g \rangle$ with and without gene duplication for noise
parameter $d = 0.0$, $0.2$, $0.4$, and $0.6$ (from top to bottom).
With gene duplication the
initial rate of gene adaptation is faster then in the single gene case.
However, at longer times the single gene case results in genes that are
closer to the target sequence $\mathcal{T}$. Other parameters are the
same as given in Fig.~\ref{fig:gene_num}.}
\end{center}
\end{figure}

Gene duplication enhances the effect of point mutation by amplifying
its effect. While the effect of a point mutation on a single gene may
relatively small when compared to the noise $\eta$, when that point
mutation is duplicated numerous times so is it's effect on the fitness
of the organism. This can mean the difference between being a mutation
that is effectively washed out by noise or strongly selected.
Fig.~\ref{fig:gene_fit_noise} shows that the relative speed-up from gene
duplication becomes more dramatic as the magnitude of the noise increases.
Notice how the point at which the single gene case crosses over the gene
duplication case shifts further and further to the right with increasing
$d$.

These results support the proposed mechanism for evolution of novel
proteins proposed by Bergthorsson \textit{et
al}~\cite{bergthorsson2007osd}. In particular, the primary steps of
gene duplication to enhance expression, followed by the slower mutation
and selection of gene with better catalytic properties then the
original, and finally reduction of gene copies all appear to have been
captured by this simple model. Note that in the long time limits we
tested, the average gene number $\langle n \rangle$ remained above
unity, probably reflecting the fact that there is a low probability of
simultaneous beneficial mutations that would allow for a favorable gene
deletion. Also, the standard deviations in Figs.~\ref{fig:gene_num},
~\ref{fig:org_fit}, and~\ref{fig:gene_fit} hint at the important role
of population variance in adaptation~\cite{boettigera2010fluctuation,zhang2010change}.
In particular, notice that the variance in gene number and in gene
fitness in Figs.~\ref{fig:gene_num} and~\ref{fig:gene_fit}, respectively,
both play a role in the additional population variance seen in
Fig.~\ref{fig:org_fit}. This results in a organismal
fitness variance that is greater in the case with gene duplication than
without despite a larger gene variance in the single gene case.

\section{Parameter Choice}

In order to interpret these simulation results in terms of their impact on
biological evolution, it is important to consider how to map the parameters
from simulation to those of real biology. A direct mapping where simulation
models biological processes at biological rates would be one particular
solution. However, the microbial world is one of very large numbers, making
this approach unfeasible. Thus, it is often best to turn to other approaches
such as scaling analysis in order to estimate how a system will behave at
a parameter setting that is far from computationally tractable.

The parameters chosen for the simulations described above are from matching
those of real biology. Realistically, a microbe contains around 1000 genes, and
the rates of mutation ($10^{-9}$) and gene duplication ($1$) given by Ref.~\cite{bergthorsson2007osd} describe per organism rates per generation. Our
simulation is intended to model a particular slice of the genome, representing
how one particular gene within the cell might be subject to particular conditions
and subsequently gene family expansion and contraction within a population. Thus,
the biologically relevant regime would be a mutation rate of $10^{-12}$ with
duplication rates of $10^{-3}$. This is fairly far from our choice of $10^{-2}$
and $1$ for these two parameters, respectively. The rates of these basic parameters
influence the timescales within the simulations. Thus we can expect that realistic
parameter range simulations would require at least 10 orders of magnitude longer
simulation times based on the mutation rates alone. Even then, these simulations
are still cannot be considered realistic in light of microbial populations that
easily number in the billions.

\begin{figure}[t]
\begin{center}
\includegraphics[width=0.9\columnwidth]{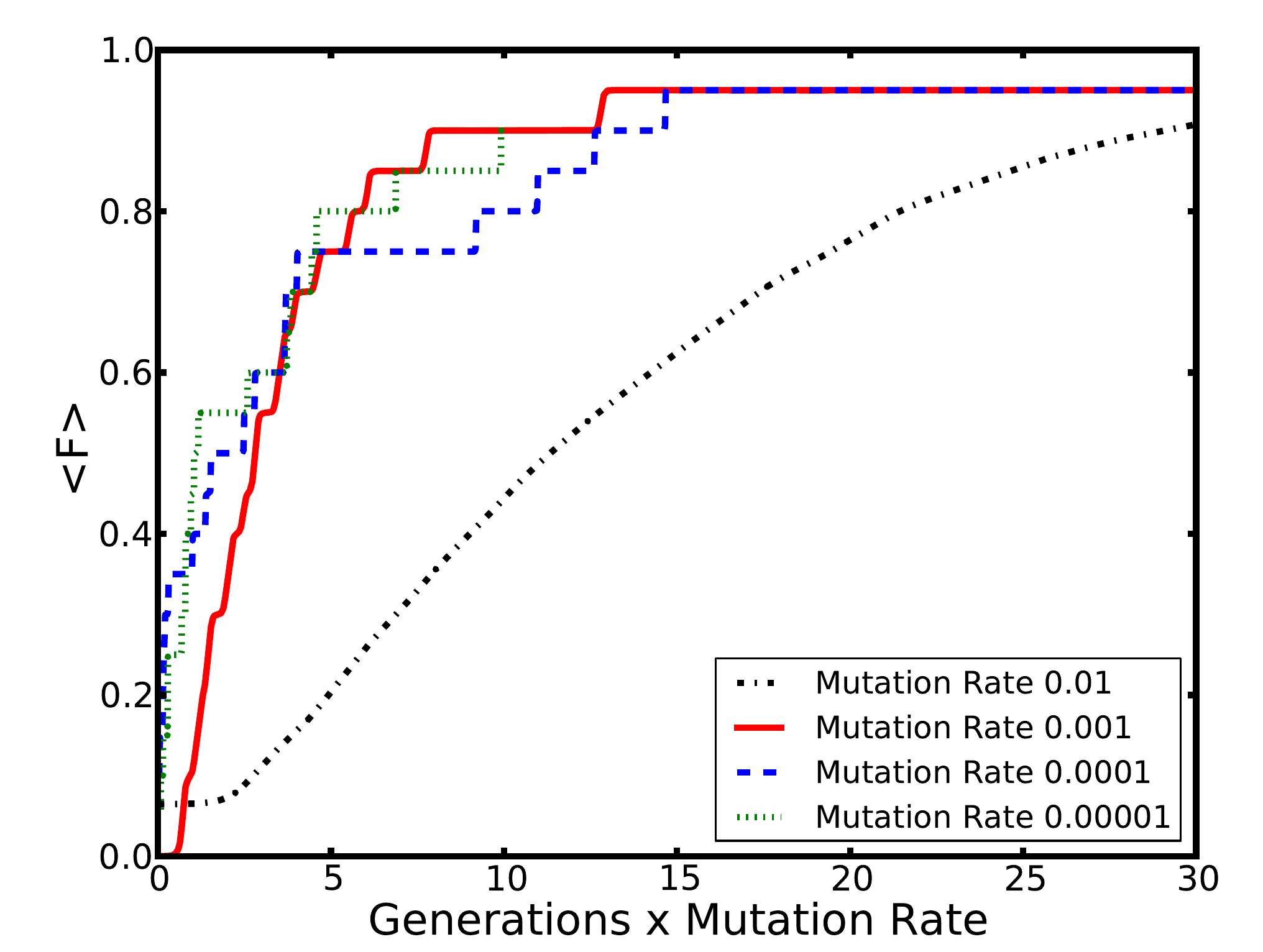}
\caption{\label{fig:no-dupl-rescaled} (Color online).  Rescaled plot of organismal fitness
for different mutation rates without gene duplication. Average organismal fitness $\langle F \rangle$
across 100 simulations plotted against a rescaled x-axis of generations times mutation rate. Other
parameters are the same as given in Fig.~\ref{fig:gene_num}. This plot shows that the rate of organismal
adaptation scales in proportion to the rate at which point mutations are produced.}
\end{center}
\end{figure}

\begin{figure}[t]
\begin{center}
\includegraphics[width=0.9\columnwidth]{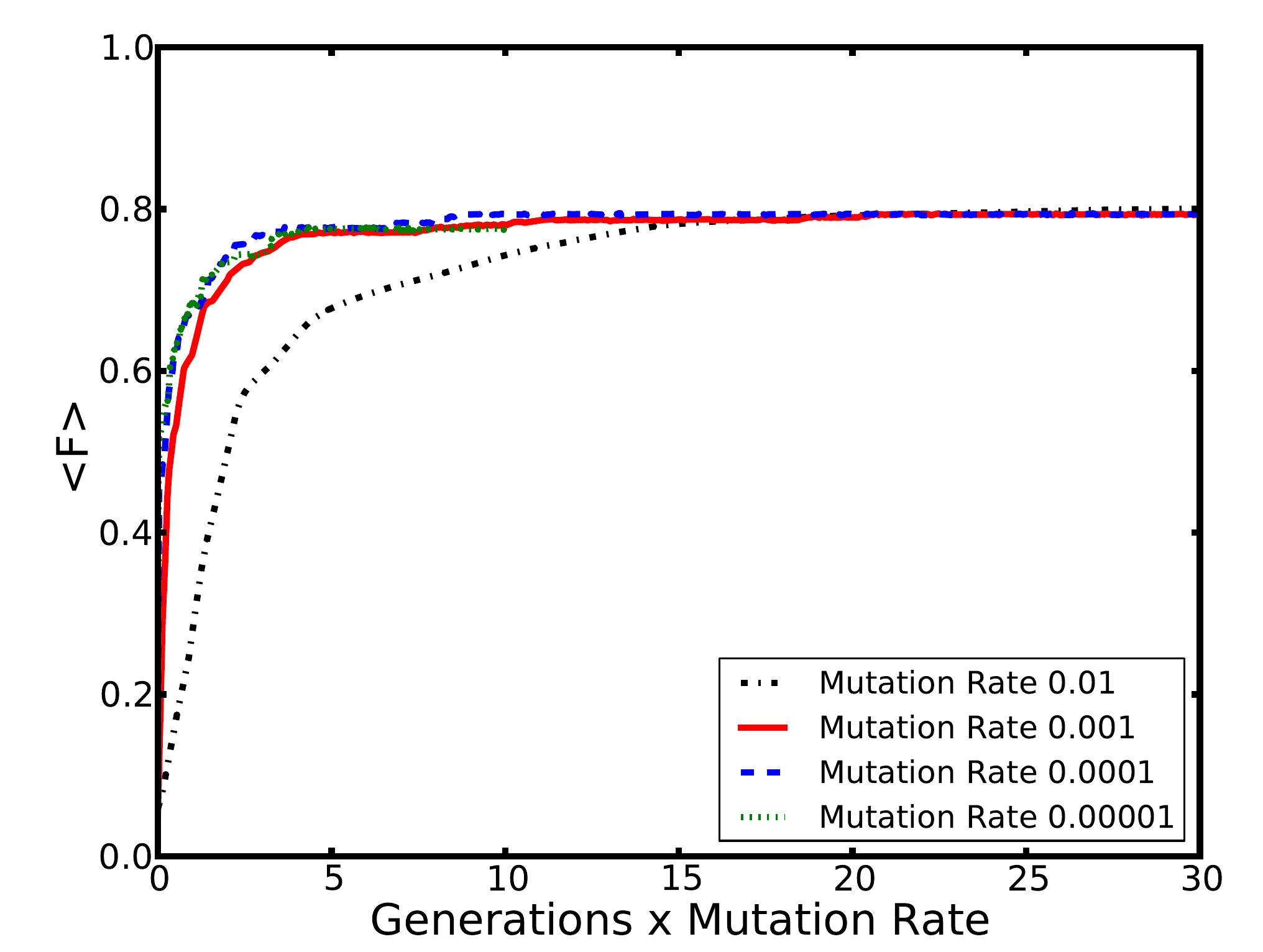}
\caption{\label{fig:with-dupl-rescaled} (Color online).  Rescaled plot of organismal fitness
for different mutation rates with gene duplication. Average organismal fitness $\langle F \rangle$ plotted against a rescaled x-axis of generations times mutation rate. Other parameters are the same as given in Fig.~\ref{fig:gene_num}. This plot shows that the rate of organismal adaptation scales in proportion to the rate at which point mutations are produced.}
\end{center}
\end{figure}

Since simulations of such size are not reasonably feasible, we instead focus on
understanding how the behavior of the system scales. In particular, mutation rate
and system size differ dramatically from biologically relevant parameters, so we
will focus our scaling analysis on these two parameters. Figs.~\ref{fig:no-dupl-rescaled} and~\ref{fig:with-dupl-rescaled} show that the rate of adaptation scales in proportion
to the mutation rate. This verifies that our above results qualitatively represent
those we would obtain for realistic biological parameters with a simple multiplicative
shift being the main difference.

\begin{figure}[t]
\begin{center}
\includegraphics[width=0.9\columnwidth]{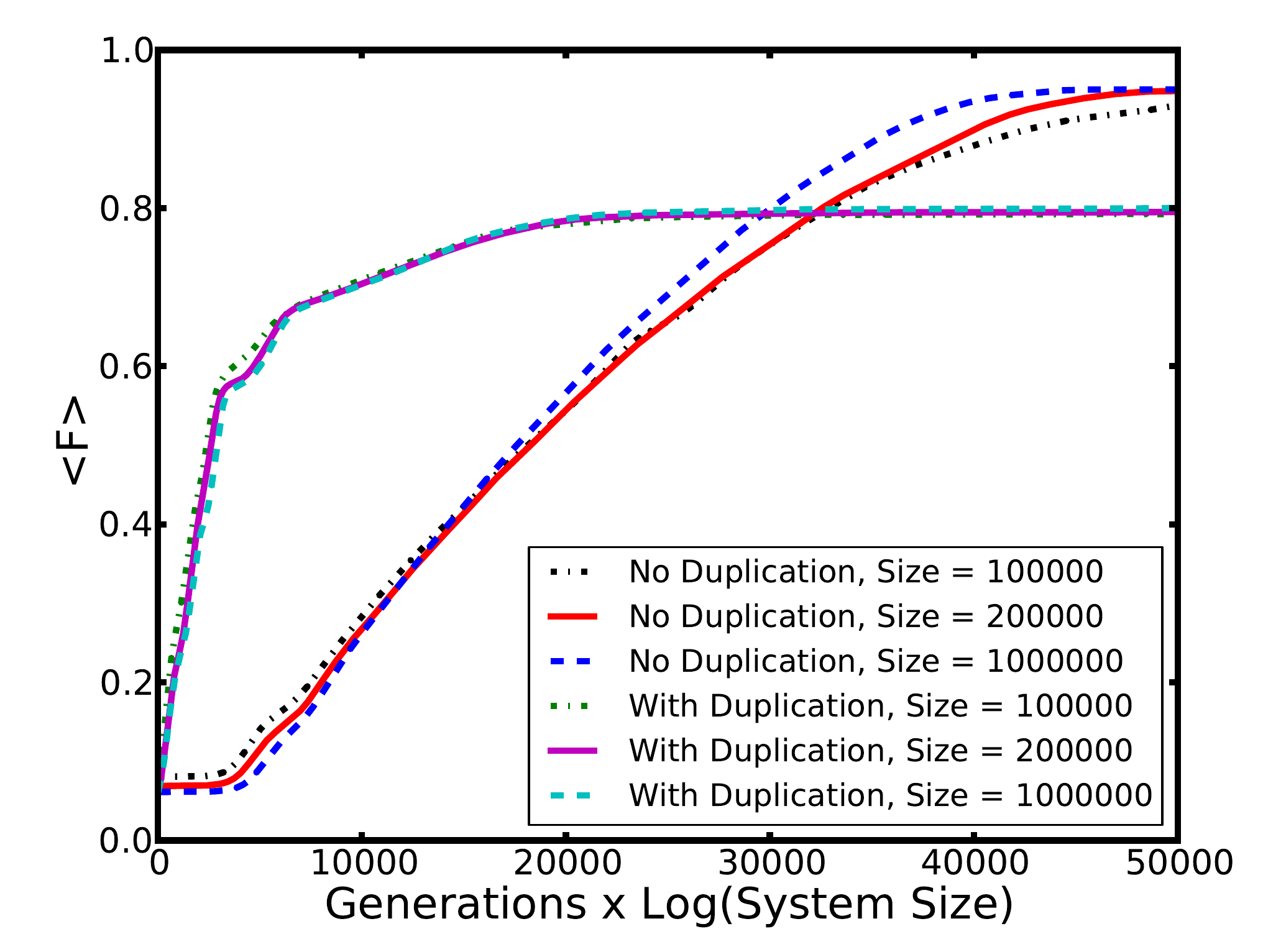}
\caption{\label{fig:size-scaling} (Color online).  Rescaled plot of organismal fitness
for different system sizes with and without gene duplication. Average organismal fitness $\langle F \rangle$ plotted
against a rescaled x-axis of generations times the natural logarithm of the system size. `No Duplication' indicates
a duplication rate of $0.0$ and `With Duplication' indicates a duplication rate of $1.0$. Mutation rate used in both
cases was $0.001$. Other parameters are the same as given in Fig.~\ref{fig:gene_num}. This plot shows that the rate
of organismal adaptation scales in proportion to the logarithm of the system size.}
\end{center}
\end{figure}

Fig.~\ref{fig:size-scaling} shows that our results scale approximately with the logarithm of the system size. Again, this scaling indicates that the speed up in adaptation presented here qualitatively hold for much larger systems, making it a realistic pathway for biological evolution.

\section{Alternate Fitness Functions}

The above model and results represent approximations to a possible biological
mechanism for generation of novel gene functions. We have attempted to keep
our model close to that of Bergthorsson {\it et al.}~\cite{bergthorsson2007osd}
in order to understand the plausibility of the gene duplication mechanism they
propose. Any study of plausibility, however, should also contain some measure of
sensitivity analysis. The sensitivity of the model to specific choice of
parameters is already discussed above. This same sensitivity check should also
be applied to our modeling choices.

In order to probe this, we put forth the following fitness function $F^2$.
\begin{center}
\begin{equation}
\mathcal{F}_k^2 = \min\left(\sum_{j=1}^n g_j^2,1\right) -
n\mu + \eta_d \label{eq:F2}
\end{equation}
\end{center}
Eq.~\ref{eq:F2} alters the linear dependence of individual gene fitnesses given
by Eq.~\ref{eq:g} to a squared dependence. This particular fitness function was
chosen because it as we increase the power to which the individual fitness $g$
is raised, we weaken the effect of the continuous selection proposed by
Bergthorsson {\it et al}~\cite{bergthorsson2007osd}. In other words, since
squaring smaller numbers reduces their value by a greater fraction than for
larger numbers closer to 1, we have are modeling a weaker initial catalytic
``side-effect'' than in Eq.~\ref{eq:F1}. Indeed,
sometimes, the benefit of the initial genes outweighs their associated fitness
costs, leading to a number of simulations that never improve in fitness.
Table~\ref{tab:yesno} shows the number of cases out of 1,000 that reach
the target sequence by the end of the run.

\begin{table}
\begin{tabular}{l|l|l}
Initial Gene & With & Without\\
Number & Duplication & Duplication\\
\hline
1 &  418 & 1000\\
2 &  858 &  956\\
3 &  991 &  977\\
4 & 1000 &  902\\
5 & 1000 &  888\\
10& 1000 &    0
\end{tabular}
\caption{Summary of number of simulation runs with fitness function $\mathcal{F}^2$
that reach the target sequence. In order
to compute this, the average gene fitness was taken at the end of each run (10,000
generations) and were considered to have evolved sufficiently toward the target
sequence if $\langle g \rangle > 0.5$. The genomes are initialized identically, with
each initial gene being chosen at random. As the initial number of genes grows, the
probability of a single gene being of sufficient benefit to outweigh it's cost
increases. Thus, the probability of success with gene duplication increases with
initial gene number accordingly. However, in the case without duplication, since
these genomes can neither duplicate nor delete genes, the extra initial genes
only provide an extra fitness cost.  Over 99\% of the cases were
$\langle g \rangle > 0.9$ or $\langle g \rangle < 0.01$.}
\label{tab:yesno}
\end{table}

\begin{figure}[t]
\begin{center}
\includegraphics[width=0.9\columnwidth]{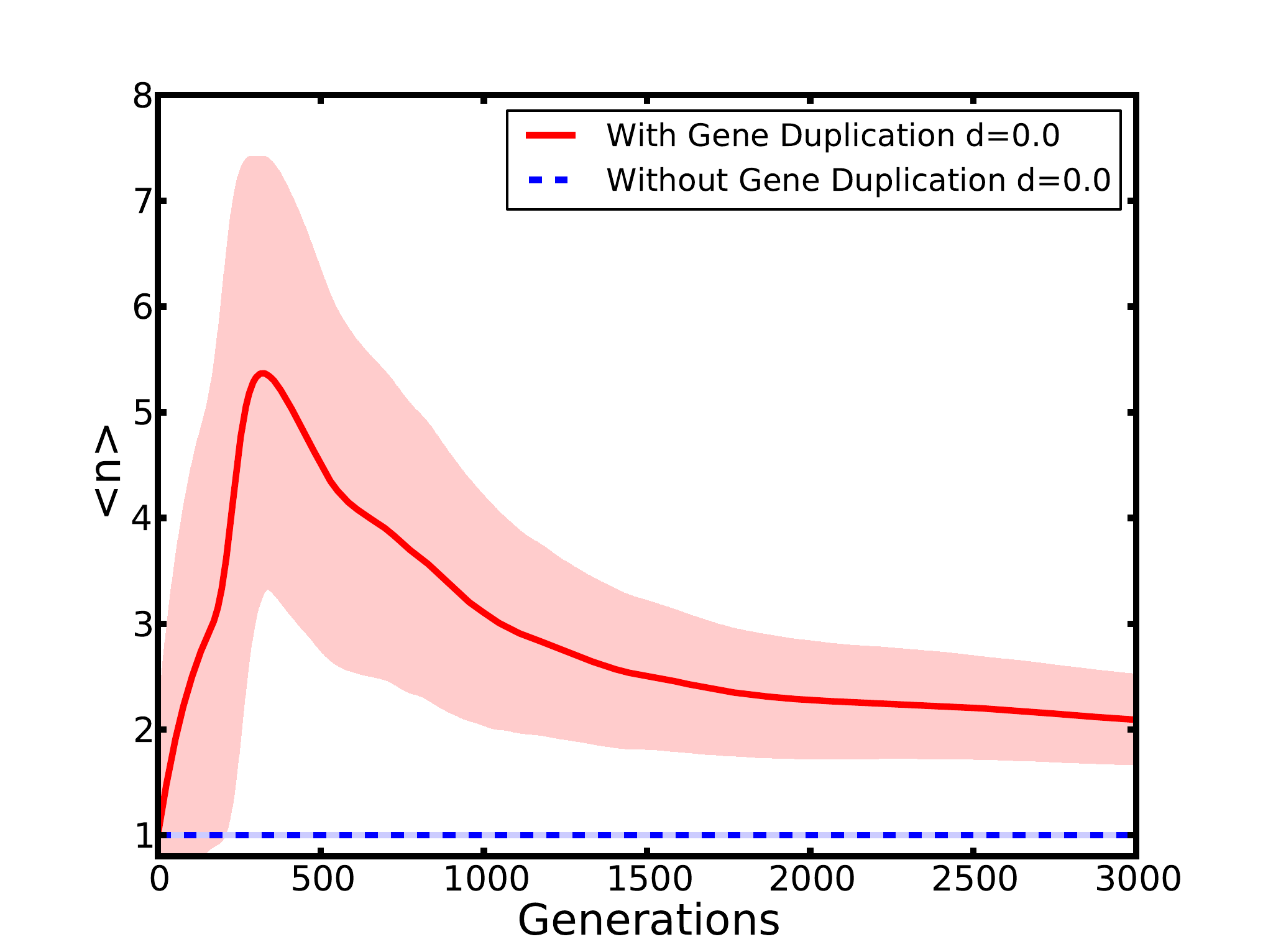}
\caption{\label{fig:gene_num2} (Color online). Average gene number
$\langle n \rangle$ with and without gene duplication for simulations
with fitness function given by Eq.~\ref{eq:F2}. Average was taken
over the 418 and 1000 simulation runs that adapted to the target
sequence with and without gene duplication, respectively. In the initial
phases of adaptation, gene duplication dominates as the primary mode
for enhancing average fitness. As time passes, the slower mode of
adaptation provided by sequence mutation refines the genes and the
average number of genes per genome decreases. In the case with no gene
duplication or deletion the gene number remains constant. Parameters
not mentioned above are the same as given in Fig.~\ref{fig:gene_num}.}
\end{center}
\end{figure}

\begin{figure}[t]
\begin{center}
\includegraphics[width=0.9\columnwidth]{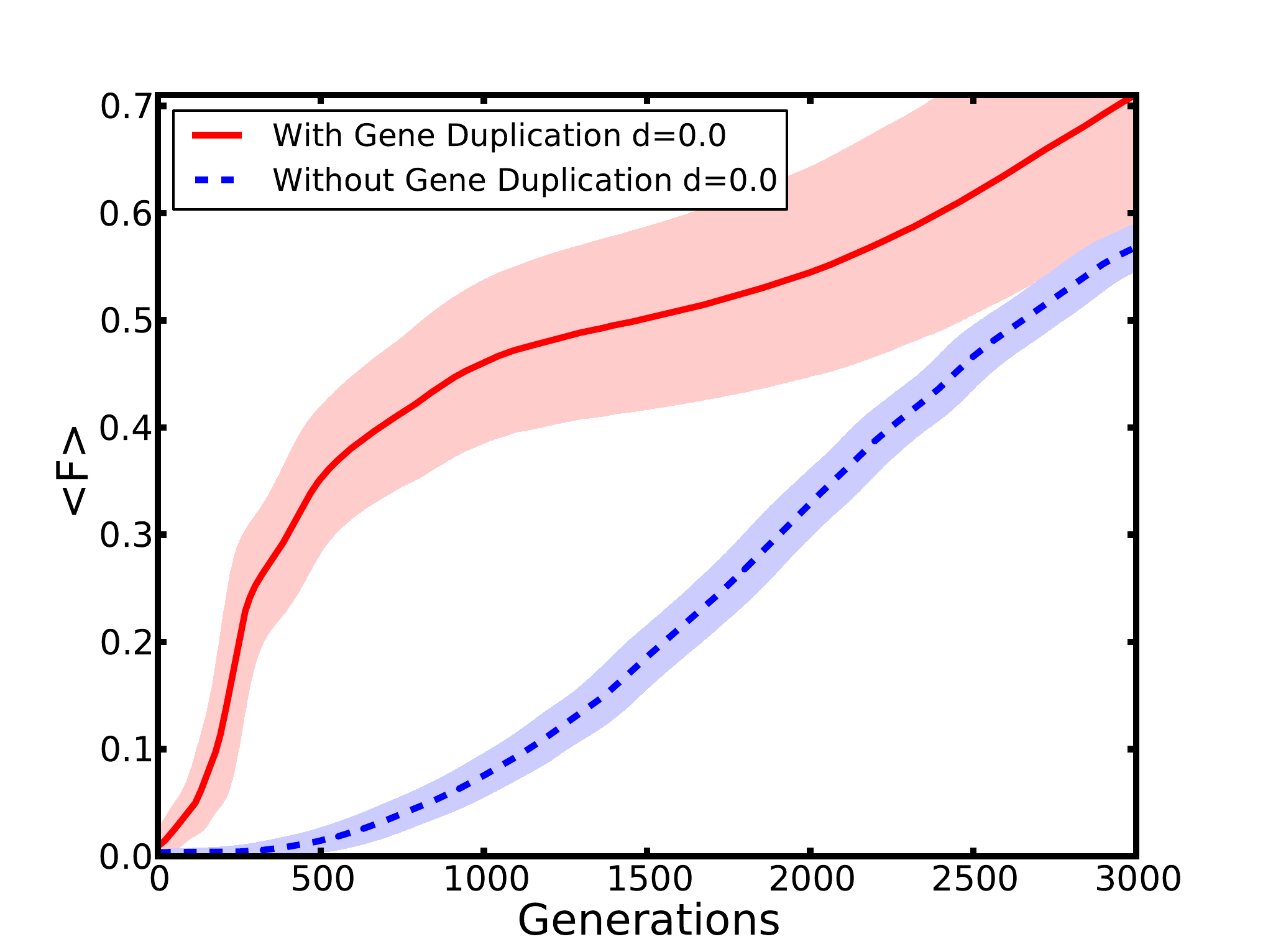}
\caption{\label{fig:org_fit2} (Color online).  Average organismal fitness
$\langle \mathcal{F}^2 \rangle$ with and without gene duplication. Average
was taken over the 418 and 1000 simulation runs that adapted to the target
sequence with and without gene duplication, respectively. Adaptation with
gene duplication is faster then in the single gene case. However, at long
times the single gene case catches up to the gene duplication case.
Parameters are the same as given in Fig.~\ref{fig:gene_num}.}
\end{center}
\end{figure}

\begin{figure}[t]
\begin{center}
\includegraphics[width=0.9\columnwidth]{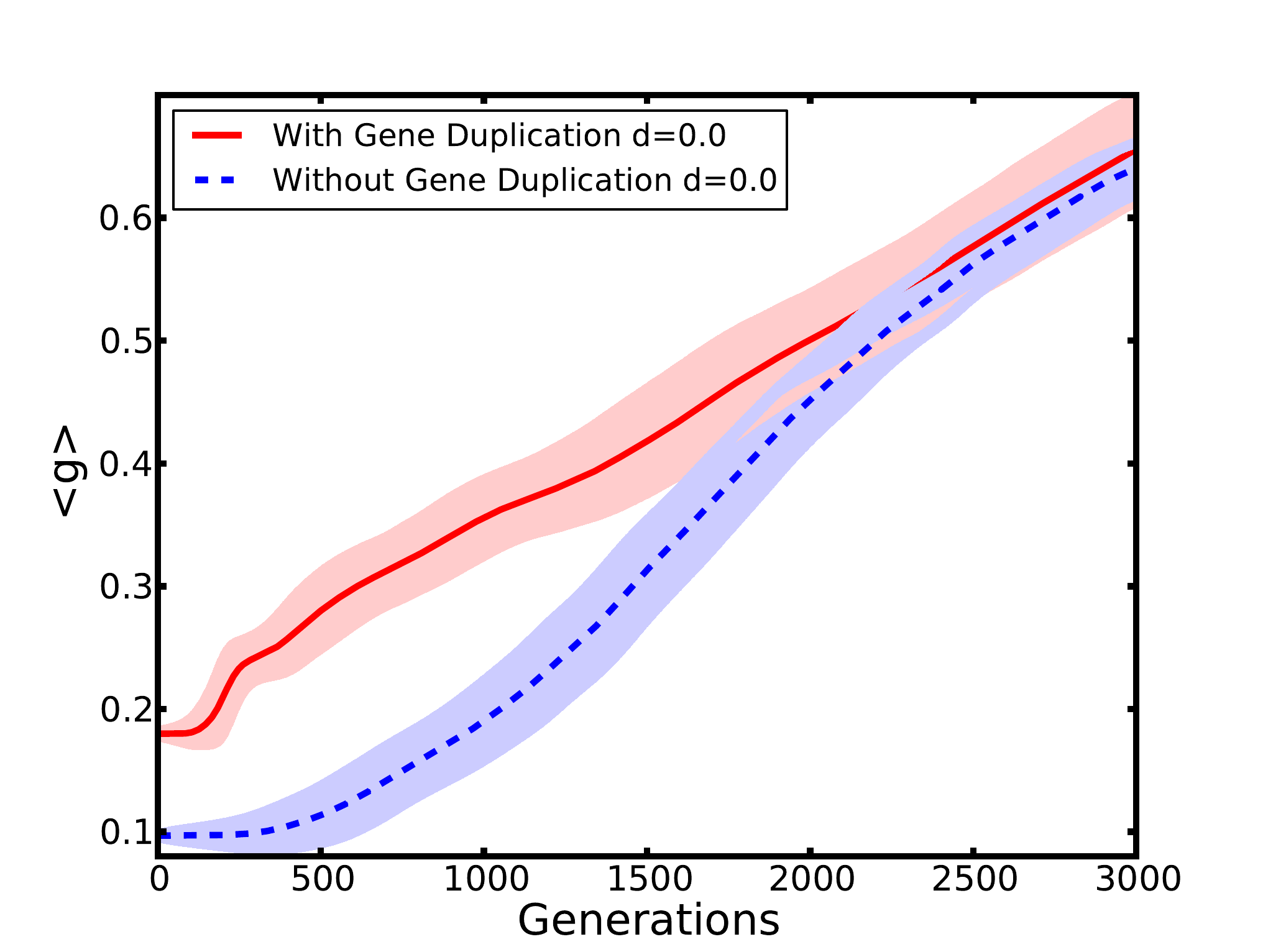}
\caption{\label{fig:gene_fit2} (Color online).  Average gene fitness
$\langle g \rangle$ with and without gene duplication for simulations
with the fitness function given by Eq.~\ref{eq:F2}. Average was taken
over the 418 and 1000 simulation runs that adapted to the target
sequence with and without gene duplication, respectively. Note that
the gene duplication scheme requires an initial gene with greater
fitness in order to adapt toward the target sequence $\mathcal{T}$.
With gene duplication the initial rate of gene adaptation is faster
than in the single gene case. Despite the initial advantage present
in the gene duplication case, at longer times $\langle g \rangle$
for the single gene case reaches similar levels.
Parameters are the same as given in Fig.~\ref{fig:gene_num}.}
\end{center}
\end{figure}

Figs.~\ref{fig:gene_num2},~\ref{fig:org_fit2}, and~\ref{fig:gene_fit2}
show the behavior of the model under the fitness function given by
Eq.~\ref{eq:F2}. All three plots show qualitatively similar properties
to those for given in section~\ref{sec:results} for the fitness function
in Eq.~\ref{eq:F1}. Specifically, the rise and fall in gene number in the
case with gene duplication in Fig.~\ref{fig:gene_num2} and the faster
initial rises in organismal and gene fitness in Figs.~\ref{fig:org_fit2}
and~\ref{fig:gene_fit2} closely resemble the results of the previous model.

Increasing the power to which $g$ is raised in the fitness function to 3,
\begin{center}
\begin{equation}
\mathcal{F}_k^3 = \min\left(\sum_{j=1}^n g_j^3,1\right) -
n\mu + \eta_d \label{eq:F3}
\end{equation}
\end{center}
further reduces the effect of continuous selection, but again results in
qualitatively similar plots, though with a larger fraction of cases with gene
duplication that do not adapt toward the target gene (data not shown).

Ultimately, this reduction of effect continues as we raise the power
of $g$ in the fitness function
\begin{center}
\begin{equation}
\mathcal{F}_k^m = \min\left(\sum_{j=1}^n g_j^m,1\right) -
n\mu + \eta_d \label{eq:Fm}
\end{equation}
\end{center}
until the benefit from an additional gene will only outweigh the cost
for genes that exactly match the target sequence $\mathcal{T}$.
Retaining results that are qualitatively similar to those presented in
Section~\ref{sec:results} requires some form of continuous selection that
allows for fitness contributions from multiple genes, supporting the
hypothesis that these features play a key role in the evolution of novel
gene function~\cite{bergthorsson2007osd}. Removing either of
these two attributes entirely nullifies the benefits of gene duplication
(data not shown). However, our results do not depend very strongly on the
strength of either of these features, and our findings outlined in
Section~\ref{sec:results} do not appear sensitive to the specific form
of the fitness function chosen, as long as they fall within a class of
bound fitness functions that allow multiple gene contributions and some
form of continuous selection.

\section{Discussion}\label{sec:diss}

Significantly, the results for the case with gene duplication
outperformed the single gene case. Continuous selection, especially in
the absence of noise, provides a means for rapid uphill adaptation.
Given that selection pressure on any one gene is
weaker when there are multiple copies, adaptation of genes could have
arguably been slower with gene duplication. We did not observe this to
be the case and found that gene duplication has an initial advantage
over the single gene case. Note that this is for the specific case
where there is a large difference between initial and the target
sequences and not true for the case where only smaller adjustments are
required to reach the target sequence. Also, we did not observe
monotonic gene duplication with a non-zero gene penalty. Beneficial
genes were able to proliferate, eventually resulting in a reduction in
the average number of genes per genome. That gene number rises and
falls according to fitness, overriding the duplication and deletion
rates for a broad range of parameters, shows us that the selective
advantage, as we represent it here, is enough to overcome these intrinsic
rates. If it were otherwise, we would not be able to posit gene amplification
as the driving dynamic behind changes in IS density.

\subsection{IS Density Following Host-Restriction}

IS elements and gene duplications go hand-in-hand with one another.
IS elements copy and paste segments of the genome from one location to
another location and thus are vehicles of gene
duplication~\cite{kleckner1981transposable,feschotte2007dna}. Enhancing
gene duplication rates allows an organism to take better advantage of
the mode of adaptation described in this work.

Host-restriction is defined as the process by which a previously
free-living organisms becomes an obligate organism, i.e., the
transition from a more independent organism to an organism dependent on
a host for survival. Genome comparisons among \textit{Buchnera}
indicate that this process involves an initial period of massive
deletions, large scale rearrangement, and the proliferation of
repetitive elements followed by extreme stability and a slow loss of
additional genes~\cite{moran2003tracing}. The pattern of repetitive
element proliferation is consistent with Fig.~\ref{fig:gene_num} in
which we see an initial spike in IS density (given by gene number $g$)
followed by a slow decrease.

\begin{figure}[t]
\begin{center}
\includegraphics[width=0.9\columnwidth]{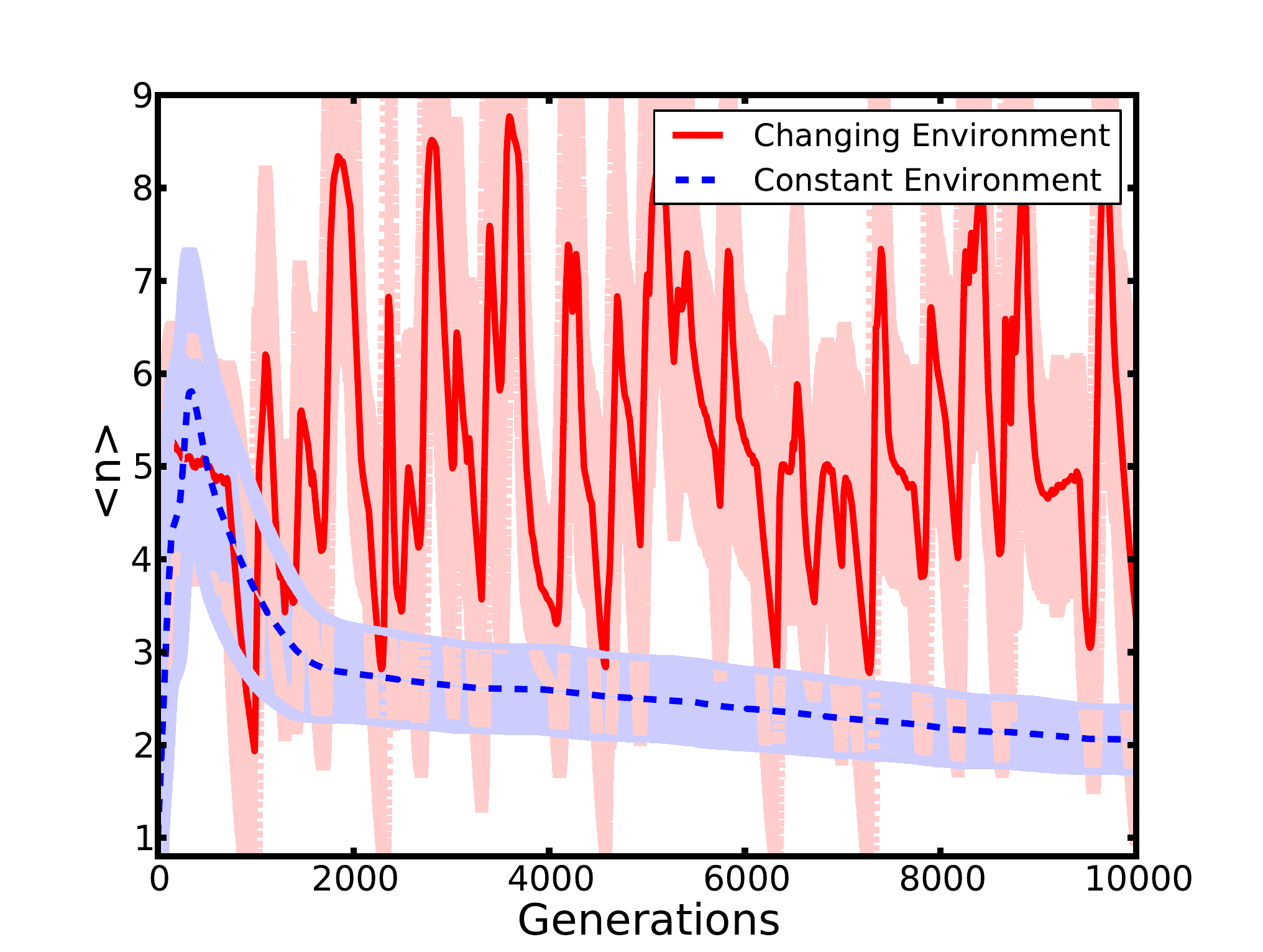}
\caption{\label{fig:free} (Color online).  Gene number
$\langle n \rangle$ with gene duplication under changing and constant
environments. The fitness function is given by Eq.~\ref{eq:F1}.
Example trajectory for changing environments was plotted by changing the
target sequence $\mathcal{T}$ to a random sequence at randomly selected
intervals according to a $0.04$ probability of change per generation.
The curve for constant environment is comprised of the averaged data
from Fig.~\ref{fig:gene_num} and is plotted here for comparison purposes.
Other parameters are the same as given in Fig.~\ref{fig:gene_num}. Notice
that the number of genes in the changing environment are generally larger
than in the case for a constant environment.}
\end{center}
\end{figure}

Wide surveys of genomes reveal that organisms with recently formed
obligate associations show an increased level of IS density in
comparison to free-living organisms~\cite{moran2004genomic}.
Conversely, ancient obligate organisms generally show a much lower
IS density in comparison to free-living
organisms~\cite{moran2004genomic}. In the context of our model, as
IS elements proliferate, they grow in number and overall
density within a genome. The pattern of boom and bust in IS
density seen in the
literature~\cite{moran2003tracing,moran2004genomic,plague2008extensive}
corresponds to a cycle of rapid adaptation to a new environment. The level
of transposon density in ancient obligate organisms is then predicted by
the long-time asymptotics of the simulations in Fig.~\ref{fig:gene_num}.
Note that we do not see a reduction from 2 genes to 1 on the timescales of
our simulation due to the fact that this requires a combination of
point mutations to occur in order to increase fitness, which then makes
the $n=2$ to $n=1$ transition very rare and slow. On the other hand,
this qualitatively matches the finding that only ancient obligate organisms
(i.e., after long times) exhibit exceedingly low transposon densities. In our
simulations, we mimicked such a rapid change by choosing random target
and starting sequences. In other words, organisms that are newly
introduced into the host environment must undergo sizable adaptive changes
in order to better compete and survive in their new conditions. The
reduced number of transposable elements and gene duplications after
long times comes from the long-time consistency that the host
environment provides. Lastly, free-living organisms face varying
environmental challenges as they migrate and their open-system
environments change. From time to time, challenges requiring rapid
changes arise and IS density increases rapidly but decreases slowly.
Our model posits that due to the occasional occurrence of these challenges,
the IS density of free-living organisms never quite falls to the
same level as that of the ancient obligate organisms. Fig.~\ref{fig:free}
shows how changes in the environment might result in an increase in the
number of transposons within a genome that is dependent on the frequency
and magnitude of these changes. Fig.~\ref{fig:free} shows the case of many
rapid and dramatic changes to environment in order to better highlight
this dynamic and differentiate it from the case with constant environment.
Free living organisms may generally undergo fewer of these changes or
changes of smaller magnitude than the initial change that occurs when an
organism initially enters a new host. This may make sense given that the
host environment introduces many new factors including immune factors that
free-living organisms have not dealt with previously.

\subsection{The Role of Noise}\label{sec:noise}

The additive noise present in the fitness function given by
Eq.~\eqref{eq:F1} plays an important role in determining the relative
advantage of gene amplification over single gene evolution. In other
words, noise determines how advantageous gene duplication will be for a
population. The greater the noise, the more difficult the evolutionary
problem of finding the optima becomes. At the same time, these more
difficult evolutionary problems are particularly suited for gene
duplication.

Consider for a moment the role and source of noise. Noise represents
the coupling between the genome and the organismal fitness---the more
noise, the weaker the coupling, and vice-versa. In principle, noise can
arise from several sources. For instance, environmental fluctuations may
destroy an organism and kill
without regard for the organismal phenotype. Conversely, it is possible
that every organism has an almost equal probability of reproducing at any
given timepoint (even though on average the fitter organism will still retain
a fixed advantage given by the noiseless fitness function). Nonetheless,
this change still dramatically alters the timescale on which selection acts.
Noise is thus not a proxy of environmental harshness, but
instead of the sensitivity of selection, or selectivity.

Selection favors phenotypes that grow, survive, and reproduce more
prolifically than their neighbors. If small changes in genotype result
in large changes to the survival and reproduction of the organism, then
the coupling between genotype and selection can be said to have a
stronger effect than noise. Note that the overall rates of reproduction
do not matter, but instead, competition is the dominant factor. Selection
pressure is not a proxy for the harshness of the environment. An
environment that kills indiscriminately is just as selective as an
environment that allows indiscriminate growth within a finite capacity.

We must now differentiate between the selectivity (or conversely, the
noise) of the system from the average environmental conditions or
directionality of selection. When the environment changes on longer
time scales, the direction of selection changes. We say that an
environment is relatively stable or consistent when the average
environment remains essentially constant over time with small
fluctuations. However, these environmental fluctuations are not the
same as noise in our model, which represents small scale fluctuations
that essentially randomize an organism's probability of survival or
reproduction. Thus, we regard the host-restriction that necessitates
rapid adaptation to the host environment as different from the
selectivity of the environment.

Indeed, in our model the noise or selectivity of the system remains an
important, but separate, contributor. An alternate explanation for the
boom in IS density following host-restriction can be seen in
Fig.~\ref{fig:gene_fit_noise}. As the noise parameter $d$ increases,
the advantage of gene duplications increases. Thus, it is possible that
simply by entering a noisier environment one should see an increase in
the number of IS elements. However, this explanation cannot account for
the later decrease in IS density in the ancient obligate organisms.

\section{Conclusion}~\label{sec:con}

We have presented a model for the dynamics of a population of microbial
genomes following a change in environmental conditions.  Our results
indicate the advantages of higher IS density in accelerating
the process of adaptation to different environmental conditions, and
the ensuing decrease in IS density during subsequent
restabilization of the environment. This corroborates evidence from
observational bioinformatics\cite{moran2003tracing} that indicates
increased IS density following host-restriction---a large
environmental change. Although our discussion has primarily focused on
host-restriction, genomic surveys can track transposon number along
environmental gradients, for example as a function of depth in the
ocean\cite{delong2006community,konstantinidis2009comparative}, and our
results should be relevant to interpretation of these
data\cite{chia2010statistical}.

\begin{acknowledgments}
The authors thank Elbert Branscomb, Thomas Butler, Jim Davis, Edward Delong, Nicholas Guttenberg, Zhenyu Wang, and Carl Woese for useful discussions. This work is
partially supported by NSF grant number NSF-EF-0526747. NC also
receives support from the Institute for Genomic Biology at the
University of Illinois at Urbana-Champaign.
\end{acknowledgments}

\bibliography{bibfileD}

\end{document}